\documentclass{article}
\usepackage{spconf,amsmath,graphicx}
\usepackage{amsfonts} 
\usepackage{bm}       

\usepackage{url}
\usepackage{booktabs}
\usepackage{multirow}

\usepackage{xcolor} 

\title{Gaussian process imputation of multiple financial series}
\name{Taco de Wolff, Alejandro Cuevas, Felipe Tobar\thanks{This work was supported by Fondecyt-Iniciaci\'on 11171165 and Conicyt-PIA AFB 170001 Center for Mathematical Modeling.}}
\address{Center for Mathematical Modeling\\Universidad de Chile}

\begin{document}
\ninept

\maketitle

\begin{abstract}
In Financial Signal Processing, multiple time series such as financial indicators, stock prices and exchange rates  are strongly coupled due to their dependence on the latent state of the market and therefore they are required to be jointly analysed. We focus on learning the relationships among financial time series by modelling them through a multi-output Gaussian process (MOGP) with expressive covariance functions. Learning these market dependencies among financial series is crucial for the imputation and prediction of financial observations. The proposed model is validated experimentally on two real-world financial datasets for which their correlations across channels are analysed. We compare our model against other MOGPs and the independent Gaussian process on real financial data.
\end{abstract}


\section{Introduction}
Financial applications of artificial intelligence research has become an area of rapid development that strives to forecast financial indicators and performance metrics using machine learning (ML) with promising results~\cite{lin2012machine,akansu2016financial,spiegeleer2018machine,gonzalvez2019financial}. Traditionally, financial indicators have been modelled using ARCH or GARCH models~\cite{engle1982autoregressive,bollerslev1986generalized,bollerslev1992arch}. More recently, flexible ML-based models have been constructed with applications to financial data~\cite{han2015financial,rizvi2017novel,tong2019discovering}.

The movement of two (or more) financial indicators such as stocks and commodities in a similar fashion (also called co-move\-ment) can be caused (i) by their mutual dependence, (ii) through changes in the same external factors that influence its price (e.g.\ political announcements, natural events, etc.), or (iii) through influences from a more complex economic system~\cite{pindyck1988excess,croux2001measure,barberis2005comovement}. In any case, discovering these relationships is crucial for investors, financial experts, and for better understanding the market. The underlying economic principles can be hard to model as these systems are rather complex. Therefore, automatic discovery of those relationships through ML may be instrumental to provide novel market insights previously unknown, as well as to confirm present conjectures. This improved understanding of the interdependence among financial indicators can greatly aid financial planning for companies and policy makers alike.

By jointly modelling financial time series as multi-output Gaussian processes (MOGPs) with rich kernel functions~\cite{han2015financial,rizvi2017novel,tong2019discovering}, we aim to discover features that are inherent to the data such as quarterly or yearly patterns or business cycles. In particular, by parametrising the positive/negative correlations between two or more time-series, the interdependence among multiple financial indicators can be trained so that a variation in one time-series can predict the movement in another time-series. Given sufficient data and the availability of recurring effects (i.e.\ patterns), we expect to construct sound predictions of one time-series channel given the others.

In the rest of the paper we will first review classical and multi-output frameworks for Gaussian processes regression. Then in Section~3 we specify the multi-output spectral mixture kernel (MOSM) and related models. In Section~4 we show the application of MOGPs to two finance experiments, namely the gold, oil, NASDAQ, and USD index dataset and the currency exchanges of ten countries with respect to the USD\@. Finally, we discuss the results in Section~5.

\section{Background: Multi-output GP}
A Gaussian process (GP)~\cite{rasmussen_gaussian_2006} defines a non-parametric prior distribution over functions $f(x) \sim \mathcal{GP}(m(x), k(x,x'))$, where $m(x)$ is the mean function (usually assumed to be zero) and $k(x,x')$ is the covariance (kernel) function. GPs can be used as a generative model for functions within Bayesian inference, therefore, data can be used to compute a predictive posterior distribution of unseen values of $f(\cdot)$. The kernel function $k$ dictates the behaviour of the modelled function, such as its periodicity and smoothness, and encodes knowledge of the time series of interest via its functional form and parameters. The choice of kernel is central in the GP framework with the radial basis function the \textit{de facto} choice due to its smoothness properties~\cite{mackay1998introduction}. However, other more expressive yet more complex kernels have recently been considered that model, for example, periodicities~\cite{melkumyan2011multi,wilson_gaussian_2013,ulrich2016gaussian,parra_spectral_2017}.

Although the GP's literature both on methods and applications is broad, most of the works address the single-output scenario when only one time series is considered, that is, a function $f: \mathbb{R}^N \to \mathbb{R}$. The extension of the GP approach to multiple signals allows for jointly modelling $M$ output channels as \textbf{coupled} GPs, where the covariance function is a function $\mathcal{K}(x,x'): \mathbb{R}^N \times \mathbb{R}^{N} \to \mathbb{R}^{M \times M}$, with $M$ the number of channels, defined element-wise as $[\mathcal{K}(x,x')]_{ij} = k_{ij}(x,x')$ between channels $i$ and $j$. The key feature of a multi-output GP is to model covariation across channels in addition to the standard temporal covariation handled by single-output GPs. One of the main challenges of MOGP models is designing flexible covariance functions while requiring the covariance function to be positive definite for all values of $x$~\cite{rasmussen_gaussian_2006}. Additionally, since an MOGP model would require parametrisation of a larger number of correlations, its increased amount of hyperparameters results in an increase in local minima and thus makes training more difficult.

A recent approach to design general and meaningful cross-channel covariances for MOGPs is to construct them in the spectral domain, that is, to parametrise their (cross) power spectral densities. An alternative is to consider a mixture of Gaussians as was originally proposed by Wilson 2013~\cite{wilson_gaussian_2013} for the single channel case. Developments in the field of multi-output and spectral mixture kernels have led to a range of new covariance functions such as the SM-LMC~\cite{goovaerts_1997,wilson2014covariance}, CSM~\cite{ulrich2016gaussian}, and the MOSM~\cite{parra_spectral_2017}. The SM-LMC kernel introduces multi-output interpretations by linearly combining the channels and thus learning cross-channel covariances. These covariance are, however, restricted to have similar behaviour across the channels. A more flexible kernel is the CSM kernel which additionally models the phase differences across channels, allowing for non-symmetric covariance functions but still requiring strong correlation between channels. The MOSM kernel adds even more flexibility by introducing a time delay factor across channels that allows for delayed influences across channels to be modelled effectively.


\section{Model specification}
Let us establish the required notation. We define (single-output) GPs operating on input $x$ as
$$f(x) \sim \mathcal{GP}(m(x), k(x,x')),$$ where the mean function $m(x)$ and covariance function $k(x,x')$ between inputs $x$ and $x'$~\cite{rasmussen_gaussian_2006} are respectively defined by
\begin{align*}
    m(x)    &= \mathbb{E}[f(x)] \\
    k(x,x') &= \mathbb{E}[(f(x)-m(x))(f(x')-m(x'))] \\
            &= \operatorname{cov}(f(x),f(x')) \,.
\end{align*}
We say that a kernel is stationary if it can be expressed as $$k(x,x')=k(x-x'),$$ where for convenience of notation, we denote the input lag $\tau = x-x'$ and will usually refer to stationary kernels simply as $k(\tau)$.

Using Bochner's theorem~\cite{bochner1959lectures,stein2012interpolation} we can describe any stationary covariance function $k(\tau)$ and its spectral density $S(\omega)$ to be Fourier pairs respectively defined as~\cite{rasmussen_gaussian_2006}
\begin{align*}
    k(\tau) &= \int S(\omega) \exp\left(2\pi i \omega \cdot \tau\right) d\omega \\
    S(\omega) &= \int k(\tau) \exp\left(-2\pi i \omega \cdot \tau\right) d\tau \,.
\end{align*}
Using these Fourier pairs, we can specify (or parametrise) a kernel in the frequency space by only requiring it to be positive since Bochner's theorem guarantees that the corresponding covariance function is always positive-definite. We can subsequently use a mixture of $Q$ Gaussian radial basis functions (RBF) in frequency space with positive weights to yield the spectral mixture kernel~\cite{wilson_gaussian_2013}
\begin{equation}
    k(\tau)=\sum_{q=1}^Q w_q \exp\left(-\frac{1}{2}\tau^\top\Sigma_q\tau\right)\cos(\mu_q^\top\tau) \,,
    \label{eq:sm_igp}
\end{equation}
with $\mu_q\in\mathbb{R}^N$, $\Sigma_q=\operatorname{diag}(\sigma_1^{(q)}, \dots, \sigma_N^{(q)})$, and $w_q, \sigma^{(q)}_i\in\mathbb{R}_+$. The kernel defined in Eq.~\ref{eq:sm_igp} we refer to as the spectral mixture independent Gaussian process kernel (SM-IGP), as we will use it to model the outputs independently.

In order to extend the spectral mixture kernel into a multi-output kernel, we use Cram\'{e}r's theorem~\cite{cramer1940theory}, which is the multivariate extension of Bochner's theorem, to obtain the multi-output spectral mixture kernel (MOSM) as proposed by \cite{parra_spectral_2017}. The MOSM kernel between channels $i$ and $j$ at input lag $\tau$ is defined as
\begin{align}
    k_{ij}(\tau)&=\sum_{q=1}^Q \alpha^{(q)}_{ij}\exp\left(-\frac{1}{2}\left(\tau+\theta^{(q)}_{ij}\right)^\top\Sigma^{(q)}_{ij}\left(\tau+\theta^{(q)}_{ij}\right)\right) \nonumber\\
    &\quad\cdot\cos\left(\left(\tau + \theta^{(q)}_{ij}\right)^\top\mu^{(q)}_{ij}+\phi^{(q)}_{ij}\right)\,,
    \label{eq:mosm_kernel}
\end{align}
with the cross-spectral parameters defined by $\alpha^{(q)}_{ij}$ the magnitude, $\mu^{(q)}_{ij}$ the mean, $\Sigma^{(q)}_{ij}$ the covariance, $\theta^{(q)}_{ij}$ the delay, and $\phi^{(q)}_{ij}$ the phase. The channels are defined by indices $i$ and $j$. For a detailed derivation of the MOSM kernel see \cite{parra_spectral_2017}.

\begin{table*}[t]
    \centering
    \begin{tabular}{l c c c c c}
    \toprule
    Model & \multicolumn{5}{c}{Parametric relation with MOSM}\\
    \midrule
    SM-IGP~\cite{wilson_gaussian_2013} & $\alpha^{(q)}_{ij}=\omega_i \delta_{ij}$ & $\mu^{(q)}_{ij}=\mu_q$ & $\Sigma^{(q)}_{ij}=\Sigma_q$ & $\theta^{(q)}_{ij}=0$ & $\phi^{(q)}_{ij}=0$\\
    SM-LMC~\cite{goovaerts_1997,wilson2014covariance} & $\alpha^{(q)}_{ij}=\omega^{(q)}_{ij}$ & $\mu^{(q)}_{ij}=\mu_q$ & $\Sigma^{(q)}_{ij}=\Sigma_q$ & $\theta^{(q)}_{ij}=0$ & $\phi^{(q)}_{ij}=0$\\
    CSM~\cite{ulrich2016gaussian}    & $\alpha^{(q)}_{ij}=\sqrt{\omega^{(q)}_{ij}}$ & $\mu^{(q)}_{ij}=\mu_q$ & $\Sigma^{(q)}_{ij}=\Sigma_q$ & $\theta^{(q)}_{ij}=0$ & -\\
    \bottomrule
    \end{tabular}%
    \caption{MOGP kernels as particular cases of MOSM\@. Channel indices are denoted by $i,j\in\{1, \dots, M\}$, and $\delta_{ij}$ denotes the Kronecker delta between channels $i$ and $j$. The MOSM kernel is shown in eq.~\eqref{eq:mosm_kernel} and the SM-IGP, SM-LMC, and CSM kernels can be recovered applying the above constraints.}
    \label{tab:mosm_params}
\end{table*}

MOSM can be understood as a more general kernel when compared to the SM-IGP, SM-LMC, and CSM kernels which can be obtained by constraining some of the parameters in Eq.~\eqref{eq:mosm_kernel}. This is illustrated in Table~\ref{tab:mosm_params}, where both the mean $\mu$ and covariance $\Sigma$ become channel independent, where either the delay $\theta$, phase $\phi$, or both are set to zero, and where the magnitude parameter $\alpha$ is scaled. Some MOGP kernels explicitly state an $R_q$ for every $q \in 1, \dots, Q$, which uses a weighted average of $R_q$ covariance functions per $Q$. In this paper we use $R_q=1$.

\section{MOGP for financial time series}
This section implements and validates the above mentioned kernels on multi-channel financial time series. The experiments were performed using the multi-output Gaussian process toolkit\footnote{\url{https://github.com/GAMES-UChile/mogptk}} (MOGPTK)~\cite{mogptk}, which contains a number of MOGP kernels and pre-training procedures. MOGPTK builds on GPFlow \cite{alex2016gpflow}, which is in turn is backed by TensorFlow~\cite{abadi2016tensorflow} and thus allows for automatic differentiation and the use of GPUs for computations (we used an 8GB Nvidia GeForce GTX 1080). All experiments were performed by 5 trials per trained model. Parameter initialisation for all MOGP kernels was achieved by estimating the power spectral density (PSD) of each channel using Bayesian non-parametric spectral estimation (BNSE)~\cite{tobar2018bayesian} and obtaining its peaks as the means of the spectral representation. The optimisation relied on L-BFGS-B with a maximum of $5000$ iterations.

The experiment aims are as follows: the first experiment models the correlation among gold, oil, NASDAQ, and the USD\@. The second experiment correlates ten currency exchanges with the USD\@.

\subsection{Gold, Oil, NASDAG, and USD index}
We considered the co-movement and interdependence among gold, oil, stock markets, and the USD\@. It is known that gold can be used to offset losses in other assets such as declining currencies, especially against USD depreciation~\cite{reboredo_can_2014}, and therefore are expected to correlate in some fashion. On the other hand, oil and the value of the USD are linked as the price of a barrel of oil is globally expressed in USD\@. The value of the USD has shown to behave (albeit weakly) correlated with oil, especially after the global financial crisis of 2008~\cite{breitenfellner2008crude,reboredo_modelling_2012}. Additionally, any fluctuation in the price of crude oil will affect economies and supply chains that are energy dependent~\cite{kilian2009impact,filis_dynamic_2011}. We represent these market effects through the NASDAQ Composite index as it covers a broad number of (mostly information technology) companies. Using these four financial series, which have been observed to influence one another, we can model the global underlying economic tendencies that affect these commodities and indicators.
 
We considered a dataset comprising series of gold and oil prices, the NASDAQ and the USD index (henceforth referred to as GONU)~\cite{dataset_gold,dataset_oil,dataset_nasdaq,dataset_usd}, between January 2017 and December 2018 with a weekly granularity. We detrended and log-transformed the data signals and removed regions in each channel to mimic missing data. For oil we removed observations between 2018-10-05 and 2018-12-31 as well as removing $30\%$ of all observations randomly. For gold we removed observations between 2018-07-01 and 2018-10-01. Finally, for the gold, NASDAQ and USD channels we removed $60\%$ randomly. Overall, our experiment consisted of $385$ training points and $446$ test points resulting in roughly five minutes of training time for the MOSM\@. We also set a Gaussian prior on the covariance magnitudes with the standard deviation of the hyperparameter set to the maximum value of each channel.

\begin{figure}[ht]
    \centering
    \includegraphics[width=1\linewidth]{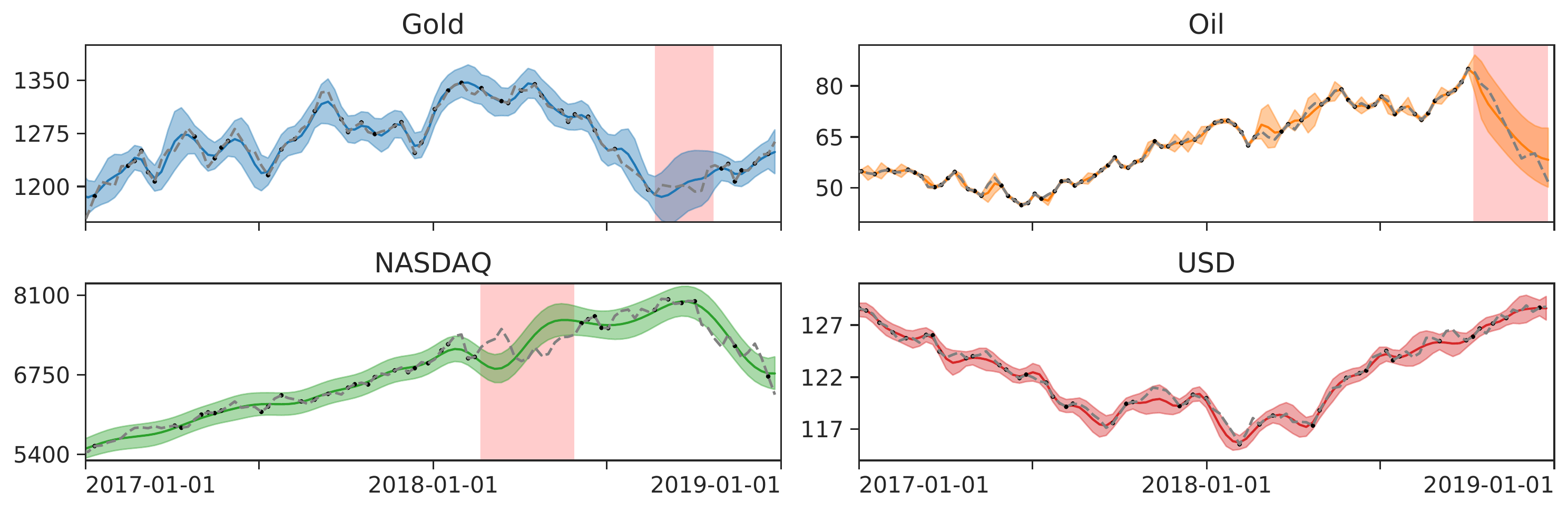}
    \caption{GONU data set with the trained MOSM kernel. Training points are shown in black, dashed lines are the ground truth and the colour coded lines are the posterior means. The coloured bands show the $95\%$ confidence intervals. The red shaded areas mark the data imputation ranges.}
    \label{fig:mosm_gonu}
\end{figure}

\begin{figure}[ht]
    \centering
    \includegraphics[width=.9\linewidth]{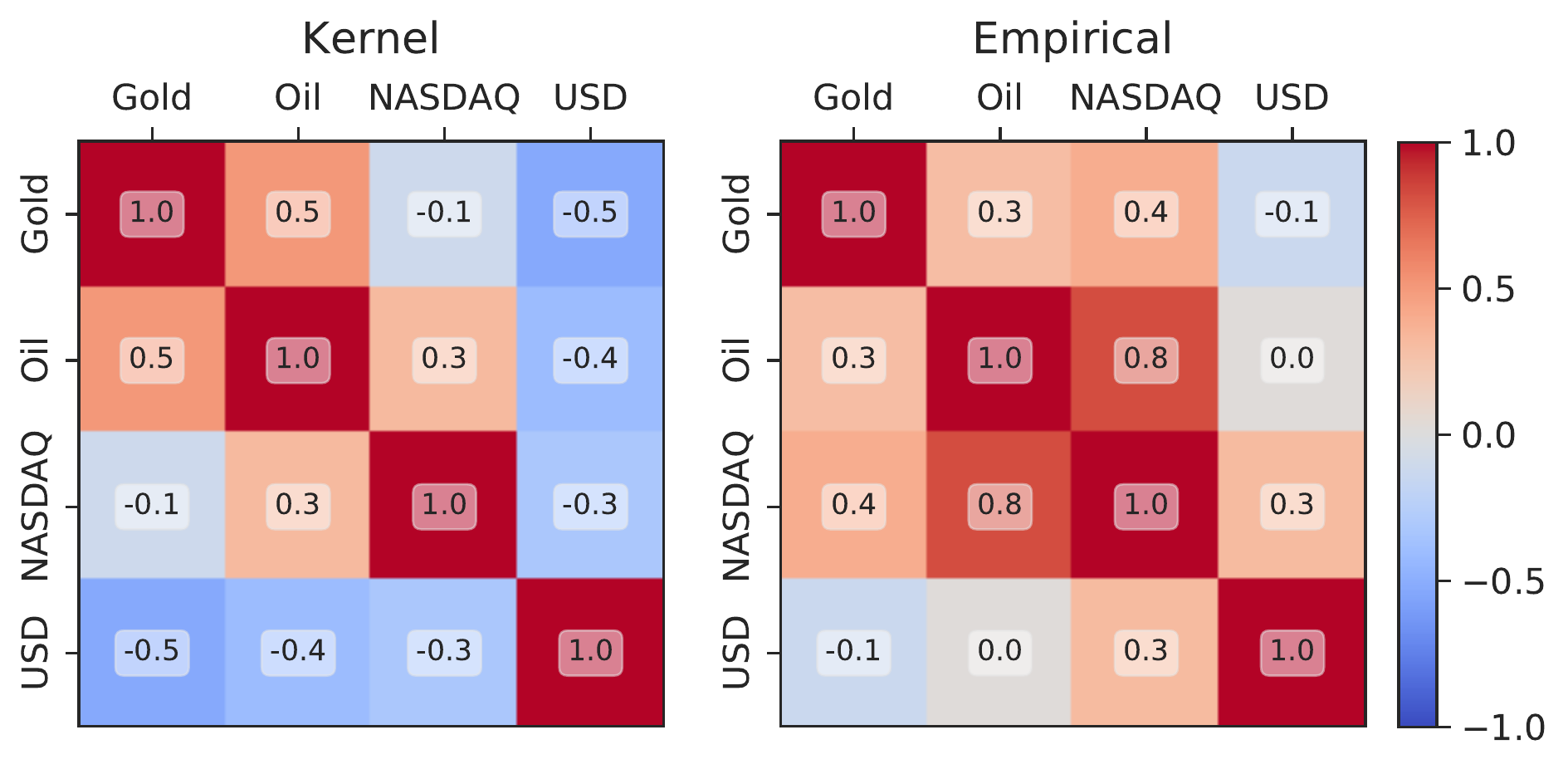}
    \caption{Cross-correlation matrix of the GONU data set (with missing data) among the channels of the trained MOSM by evaluating the (normalised) kernel (Eq.~\ref{eq:mosm_kernel}) at $\tau=0$ (left) and the empirical cross-correlation of the full data set (right). The off-diagonal elements show how much two currencies are aligned or anti-aligned, or whether they are unaligned and have negligible correlation.}
    \label{fig:mosm_gonu_correlation}
\end{figure}

Fig.~\ref{fig:mosm_gonu} shows a fit of the MOSM kernel. The MOSM model is able to encapsulate the structure of the channels with almost all data within the confidence interval of 95\%, even for parts that have missing data but with a deviating imputation for NASDAQ. The related cross-correlation matrix is plotted in Fig.~\ref{fig:mosm_gonu_correlation}. Notice that the empirical cross-correlation matrix is showing correlation between gold, oil, and NASDAQ, with especially a strong dependency between oil and NASDAQ thus confirming our hypothesis. The hedging quality of gold can also be seen (albeit faintly) with the negative cross-correlation between gold and the USD index.

Our trained MOSM kernel is recovering the more significant dependencies such as the oil and gold correlation and the oil and NASDAQ correlation. In Fig.~\ref{fig:mosm_gonu} these curves follow similar behaviour, especially for oil and the NASDAQ this is apparent. The USD is found to correlate more negatively with the other channels, as well as gold and the NASDAQ\@. It should be noted that the MOSM finds correlations by minimising the negative log-likelihood (NLL), where if three channels correlate, the model could find correlation between the first and second, and between the second and third channels, but not necessarily between the first and third, explaining the discrepancies between kernel and empirical cross-correlations. Furthermore, the MOSM only uses part of the data, and depending on the number of parameters and training it may not find all correlations. Table~\ref{tab:errors} (left) shows error values of the test set comparing different models against the MOSM\@.

\begin{table*}[ht]
   \centering
    \begin{tabular}{l c c c c}
        \toprule
        & \multicolumn{2}{c}{Gold, Oil, NASDAQ, USD index} & \multicolumn{2}{c}{Currency exchange rates}\\
        Model & nMAE ($10^{-2}$) & nRMSE ($10^{-2}$) & nMAE ($10^{-3}$) & nRMSE ($10^{-3}$) \\
        \midrule
        SM-IGP~\cite{wilson_gaussian_2013} & $2.817 \pm 0.000$      & $5.071 \pm 0.000$          
               & $5.478 \pm 0.000$      & $7.481 \pm 0.000$ \\
        SM-LMC~\cite{goovaerts_1997,wilson2014covariance} & $2.5 \pm 0.4$          & $3.4 \pm 0.6$
               & $6.6 \pm 0.5$          & $8.9 \pm 0.6$ \\
        CSM~\cite{ulrich2016gaussian}    & $1.88 \pm  0.02$       & $\mathbf{2.44 \pm 0.06}$
               & $8 \pm 1$              & $10\pm 2$ \\
        MOSM~\cite{parra_spectral_2017}   & $\mathbf{1.8 \pm 0.1}$ & $2.6 \pm 0.4$
               & $\mathbf{4.8 \pm 0.3}$ & $\mathbf{6.5\pm 0.4}$ \\
        \bottomrule
    \end{tabular}%
    \caption{Performance indices for the GONU and exchange rate experiments using the normalised mean absolute error (nMAE) and normalised root mean square error (nRMSE) on the test data and averaged over five test trials. Both are normalised by division of the mean.}
    \label{tab:errors}
\end{table*}

\subsection{Exchange Rates}
Much like the GONU data set, the movement of exchange rates among large currencies is due to international market changes and national macro economic factors. Exchange rates are heavily influenced by inflation and interest rates, trade and economic performance. We chose ten exchange rates against the USD, namely the AUD, CAD, CHF, EUR, GBP, HKD, JPY, KRW, MXN, and NZD using a daily granularity with data ranging from 2017-01-01 to 2017-12-31. For all the channels, $30$\% of the data points have been removed randomly. All channels have the last 40 days removed except for EUR, JPY, and AUD\@. The EUR, JPY, and AUD thus act as reference channels to predict the other currency exchanges. For some channels an additional range has been removed to simulate missing data. Overall, we used $1535$ training points and $955$ test points, where each trial took roughly $60$ minutes per trial for the MOSM\@.

\begin{figure}[ht]
    \centering
    \includegraphics[width=1\linewidth]{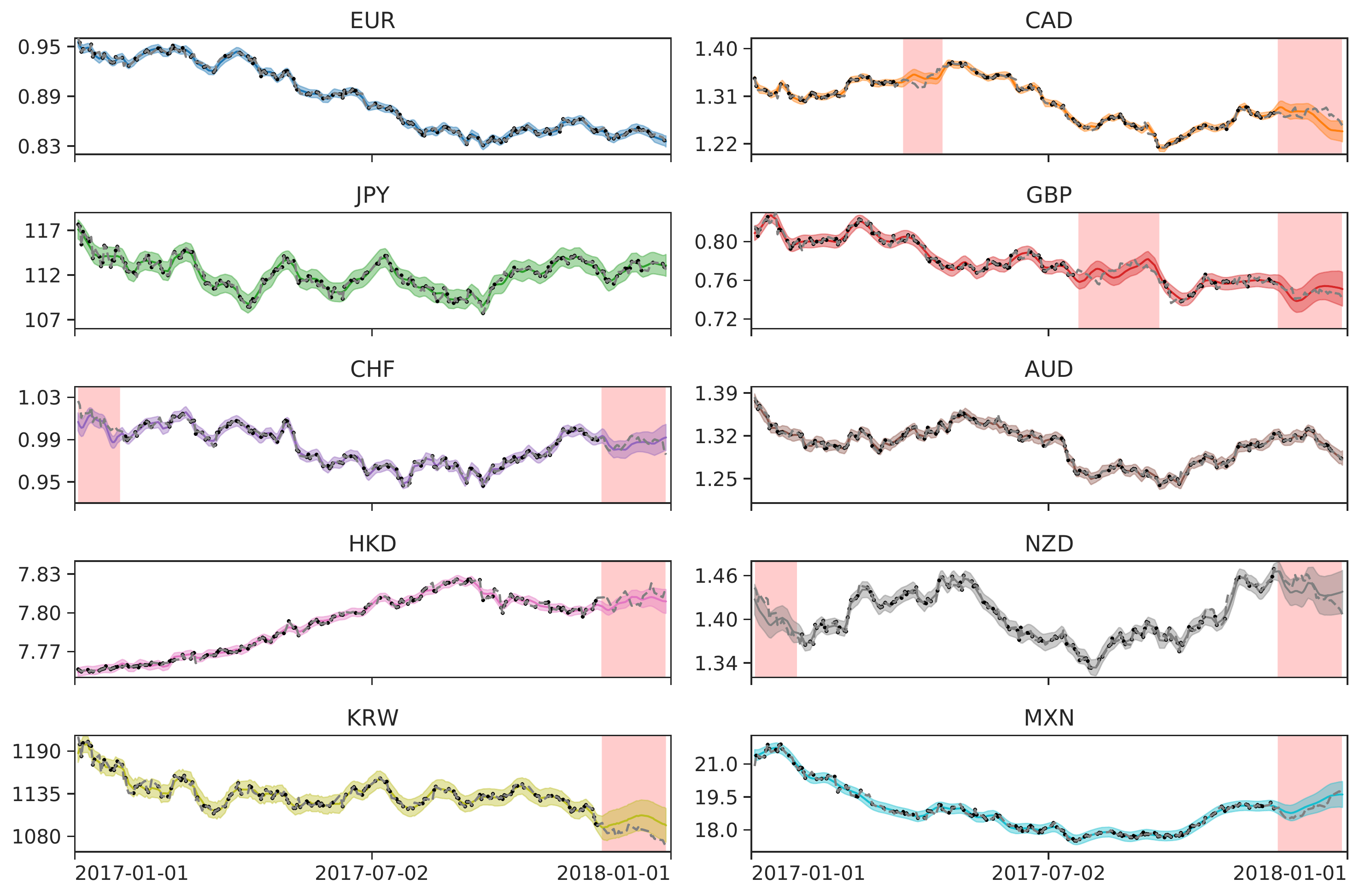}
    \caption{Ten currency exchange rates with respect to the USD fitted using the MOSM kernel. Training points are shown in black, ground truth in dashed grey, the coloured lines are the posterior means and the coloured shadows are the $95\%$ confidence intervals. The red shaded areas mark the data imputation ranges.}
    \label{fig:mosm_exchange_rate}
\end{figure}

Fig.~\ref{fig:mosm_exchange_rate} shows the currency exchange data set with a fit of the MOSM kernel. We see that the predicted posterior means at the removed tails follow the data quite closely. A possible reason why one channel can recover missing data better while other channels have difficulty doing so, lies in the fact that a strongly correlating channel is needed to impute the data. Notice that since the MOSM is a covariance-driven model, the EUR, JPY, and AUD channels can be used to reconstruct the other channels.

\begin{figure}[ht]
    \centering
    \includegraphics[width=.8\linewidth]{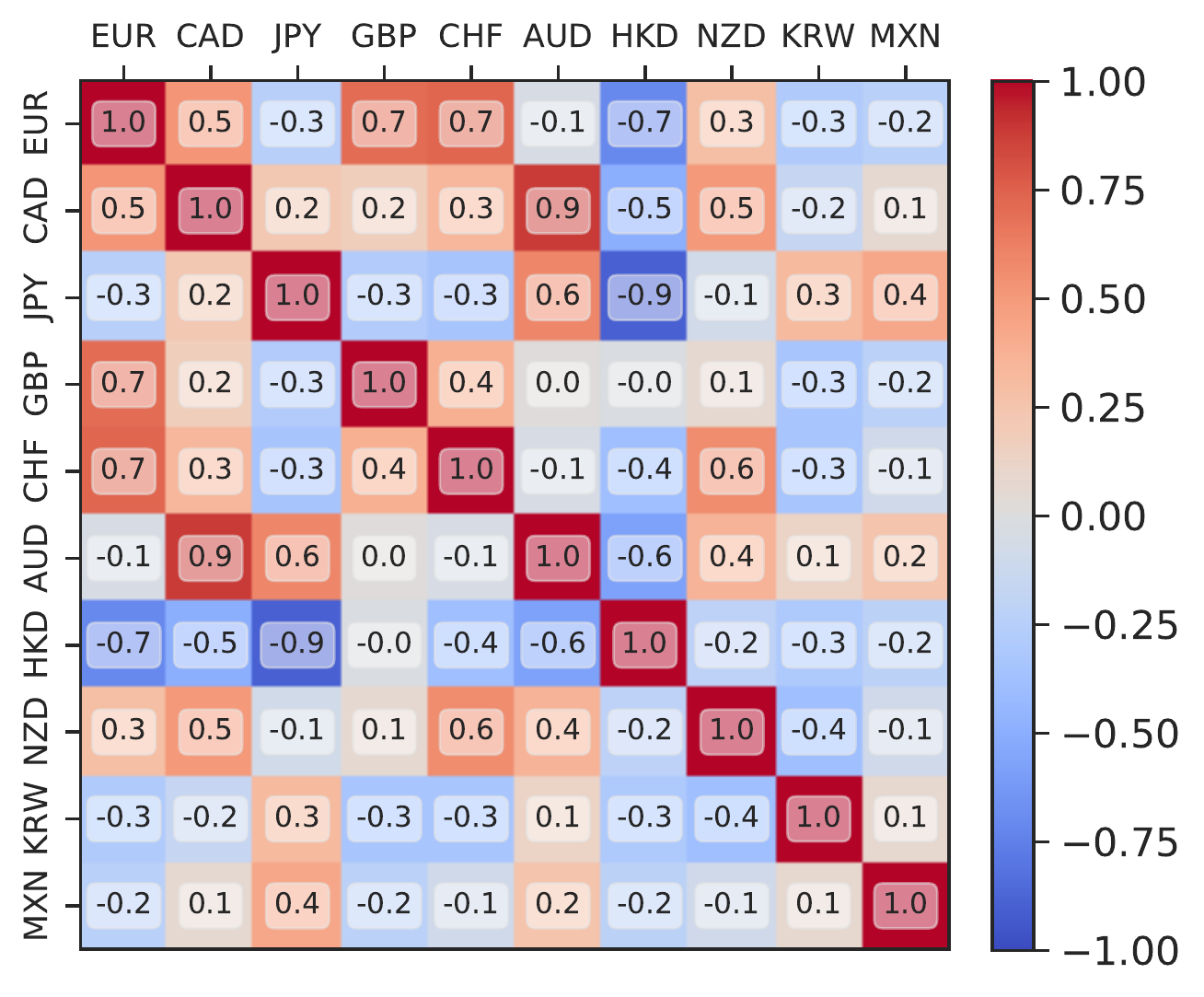}
    \caption{Cross-correlation between the ten currency exchange channels using the MOSM by evaluating the kernel (Eq.~\ref{eq:mosm_kernel}) at $\tau=0$ and normalising with the sum of the weights of each channel.}
    \label{fig:mosm_exchange_rate_correlation}
\end{figure}

Fig.~\ref{fig:mosm_exchange_rate_correlation} shows how much the channels correlate among each other under the trained MOSM kernel. Among the EUR, GBP, and CHF channels we see a strong positive correlation which is highly likely as the EU is the major trading partner for the GBP and CHF\@. Furthermore, we see that the HKD correlates negatively with the EUR, JPY, and AUD as the AUD and JPY correlate positively. The correlation between AUD and NZD is hardly surprising as these markets usually move quite similarly due to the geographic constraints of New Zealand.


\section{Discussion}
We have presented and implemented the MOGP approach through analysis of real-world financial time series. In particular, we have compared the performance of five trials of the MOSM, CSM, SM-IGP, and SM-LMC multi-output GP kernels, where we find that we are able to use the added flexibility of the MOSM to our advantage. A summary of kernel performance with respect to the normalised mean absolute error (nMAE) and normalised root mean square error (nRMSE) in the test points is given in Table~\ref{tab:errors}, where we observe a general decrease in error for models that are more flexible. The MOSM shows lower error values although it is also the most difficult model to train due the number of extra parameters. With an appropriate choice of initialisation parameters it is, however, able to find better fits between the channels than other models in terms of nMAE and nRMSE.

The challenge of fitting volatile financial data is the fact that unpredictable pattern deviations occur without precedent. While for example the GARCH model allows for modelling the heteroskedastic nature of financial data (i.e.\ the varying magnitude of volatility over time), the spectral kernels do not as they are by definition stationary which is also one of their drawbacks. While we can extract some of the interdependencies between the channels, these cross-correlations are hard to train and prone to fluctuations between trials.

Future work could include exploring financial data sets with non-Gaussian likelihoods by warping GPs as proposed by \cite{rios_tobar2019,warped04}, or by using Student's t-distribution likelihoods to better identify heteroskedasticity as used by GARCH and other financial models. Furthermore, better initialisation of hyperparameters and training can also greatly improve the results of the models which should remain an active area of research. However, the possibility of MOGPs to explore relations across channels could become a valuable asset in financial modelling and market dependency assessment.

\bibliographystyle{IEEEbib}
\bibliography{refs}

\begin{thebibliography}{10}

\bibitem{lin2012machine}
W.~{Lin}, Y.~{Hu}, and C.~{Tsai},
\newblock ``Machine learning in financial crisis prediction: A survey,''
\newblock {\em IEEE Transactions on Systems, Man, and Cybernetics, Part C
  (Applications and Reviews)}, vol. 42, no. 4, pp. 421--436, 2012.

\bibitem{akansu2016financial}
A.N. Akansu, S.R. Kulkarni, and D.~Malioutov,
\newblock {\em Financial Signal Processing and Machine Learning},
\newblock John Wiley \& Sons, Ltd, 2016.

\bibitem{spiegeleer2018machine}
J.~de~Spiegeleer, D.~Madan, S.~Reyners, and W.~Schoutens,
\newblock ``Machine learning for quantitative finance: fast derivative pricing,
  hedging and fitting,''
\newblock {\em Quantitative Finance}, vol. 18, no. 10, pp. 1635--1643, 2018.

\bibitem{gonzalvez2019financial}
J.~Gonzalvez, E.~Lezmi, T.~Roncalli, and J.~Xu,
\newblock ``Financial applications of {G}aussian processes and {B}ayesian
  optimization,''
\newblock {\em SSRN Electronic Journal}, 2019.

\bibitem{engle1982autoregressive}
Robert~F. Engle,
\newblock ``Autoregressive conditional heteroscedasticity with estimates of the
  variance of {United Kingdom} inflation,''
\newblock {\em Econometrica}, vol. 50, no. 4, pp. 987--1007, 1982.

\bibitem{bollerslev1986generalized}
T.~Bollerslev,
\newblock ``Generalized autoregressive conditional heteroskedasticity,''
\newblock {\em Journal of econometrics}, vol. 31, no. 3, pp. 307--327, 1986.

\bibitem{bollerslev1992arch}
T.~Bollerslev, R.Y. Chou, and K.F. Kroner,
\newblock ``{ARCH} modeling in finance: A review of the theory and empirical
  evidence,''
\newblock {\em Journal of Econometrics}, vol. 52, no. 1, pp. 5 -- 59, 1992.

\bibitem{han2015financial}
J.~{Han} and X.~{Zhang},
\newblock ``Financial time series volatility analysis using gaussian process
  state-space models,''
\newblock in {\em IEEE Global Conference on Signal and Information Processing},
  2015, pp. 358--362.

\bibitem{rizvi2017novel}
S.~A.~A. {Rizvi}, S.~J. {Roberts}, M.~A. {Osborne}, and F.~{Nyikosa},
\newblock ``A novel approach to forecasting financial volatility with
  {G}aussian process envelopes,''
\newblock {\em arXiv e-prints}, p. arXiv:1705.00891, 2017.

\bibitem{tong2019discovering}
A.~Tong and J.~Choi,
\newblock ``Discovering latent covariance structures for multiple time
  series,''
\newblock {\em Proceedings of the 36th International Conference on Machine
  Learning}, vol. 97, pp. 6285--6294, 2019.

\bibitem{pindyck1988excess}
R.S. Pindyck and J.J. Rotemberg,
\newblock ``The excess co-movement of commodity prices,''
\newblock {\em Economic Journal}, vol. 100, no. 403, pp. 1173--1189, 1988.

\bibitem{croux2001measure}
C.~Croux, M.~Forni, and L.~Reichlin,
\newblock ``A measure of comovement for economic variables: Theory and
  empirics,''
\newblock {\em The Review of Economics and Statistics}, vol. 83, no. 2, pp.
  232--241, 2001.

\bibitem{barberis2005comovement}
N.~Barberis, A.~Shleifer, and J.~Wurgler,
\newblock ``Comovement,''
\newblock {\em Journal of Financial Economics}, vol. 75, no. 2, pp. 283 -- 317,
  2005.

\bibitem{rasmussen_gaussian_2006}
C.E. Rasmussen and C.K.I. Williams,
\newblock {\em Gaussian {Processes} for {Machine} {Learning}},
\newblock MIT Press, 2006.

\bibitem{mackay1998introduction}
D.J.C. MacKay,
\newblock ``Introduction to {Gaussian} processes,''
\newblock {\em NATO ASI Series F Computer and Systems Sciences}, vol. 168, pp.
  133--166, 1998.

\bibitem{melkumyan2011multi}
A.~Melkumyan and F.~Ramos,
\newblock ``Multi-kernel {Gaussian} processes,''
\newblock in {\em 22nd International Joint Conference on Artificial
  Intelligence}, 2011, vol.~2, pp. 1408--1413.

\bibitem{wilson_gaussian_2013}
A.~Wilson and R.~Adams,
\newblock ``Gaussian process kernels for pattern discovery and extrapolation,''
\newblock in {\em Proceedings of the 30th International Conference on Machine
  Learning}, 2013, pp. 1067--1075.

\bibitem{ulrich2016gaussian}
K.~Ulrich,
\newblock {\em {Gaussian} Process Kernels for Cross-Spectrum Analysis in
  Electrophysiological Time Series},
\newblock Ph.D. thesis, Duke University, 2016.

\bibitem{parra_spectral_2017}
G.~Parra and F.~Tobar,
\newblock ``Spectral mixture kernels for multi-output {Gaussian} processes,''
\newblock {\em Advances in Neural Information Processing Systems 30}, pp.
  6681--6690, 2017.

\bibitem{goovaerts_1997}
P.~Goovaerts,
\newblock {\em Geostatistics for Natural Resources Evaluation},
\newblock Oxford University Press, 1997.

\bibitem{wilson2014covariance}
A.G. Wilson,
\newblock {\em Covariance kernels for fast automatic pattern discovery and
  extrapolation with {Gaussian} processes},
\newblock Ph.D. thesis, University of Cambridge, 2014.

\bibitem{bochner1959lectures}
S.~Bochner,
\newblock {\em Lectures on Fourier Integrals. (AM-42)},
\newblock Princeton University Press, 1959.

\bibitem{stein2012interpolation}
M.L. Stein,
\newblock {\em Interpolation of spatial data: some theory for kriging},
\newblock Springer Science \& Business Media, 2012.

\bibitem{cramer1940theory}
H.~Cramer,
\newblock ``On the theory of stationary random processes,''
\newblock {\em Annals of Mathematics}, vol. 41, no. 1, pp. 215--230, 1940.

\bibitem{mogptk}
T.~de~Wolff, A.~{Cuevas}, and F.~{Tobar},
\newblock ``{MOGPTK: The Multi-Output Gaussian Process Toolkit},''
\newblock {\em arXiv e-prints}, p. arXiv:2002.03471, 2020.

\bibitem{alex2016gpflow}
A.G. de~G.~Matthews et~al.,
\newblock ``{GPflow}: A {Gaussian} process library using {TensorFlow},''
\newblock {\em Journal of Machine Learning Research}, vol. 18, no. 40, pp.
  1--6, 2017.

\bibitem{abadi2016tensorflow}
M.~Abadi et~al.,
\newblock ``{TensorFlow}: A system for large-scale machine learning,''
\newblock in {\em 12th USENIX Symposium on Operating Systems Design and
  Implementation (OSDI 16)}, 2016, pp. 265--283.

\bibitem{tobar2018bayesian}
F.~Tobar,
\newblock ``Bayesian nonparametric spectral estimation,''
\newblock {\em Advances in Neural Information Processing Systems 31}, pp.
  10127--10137, 2018.

\bibitem{reboredo_can_2014}
J.C. Reboredo and M.A. Rivera-Castro,
\newblock ``Can gold hedge and preserve value when the {US} dollar
  depreciates?,''
\newblock {\em Economic Modelling}, vol. 39, pp. 168--173, 2014.

\bibitem{breitenfellner2008crude}
A.~Breitenfellner and J.~Crespo Cuaresma,
\newblock ``Crude oil prices and the {USD}/{EUR} exchange rate,''
\newblock {\em Monetary Policy \& the Economy}, pp. 102--121, 2008.

\bibitem{reboredo_modelling_2012}
J.C. Reboredo,
\newblock ``Modelling oil price and exchange rate co-movements,''
\newblock {\em Journal of Policy Modeling}, vol. 34, no. 3, pp. 419--440, 2012.

\bibitem{kilian2009impact}
L.~Kilian and C.~Park,
\newblock ``The impact of oil price shocks on the {US} stock market,''
\newblock {\em International Economic Review}, vol. 50, no. 4, pp. 1267--1287,
  2009.

\bibitem{filis_dynamic_2011}
G.~Filis, S.~Degiannakis, and C.~Floros,
\newblock ``Dynamic correlation between stock market and oil prices: {The} case
  of oil-importing and oil-exporting countries,''
\newblock {\em International Review of Financial Analysis}, vol. 20, no. 3, pp.
  152--164, 2011.

\bibitem{dataset_gold}
``{LMBA} gold price,''
  \url{https://fred.stlouisfed.org/series/GOLD​AMGBD228NLBM},
\newblock Accessed: 2019-09-01.

\bibitem{dataset_oil}
``Brent oil price,'' \url{https://www.eia.gov/dnav/pet/hist/RBRTEd.htm},
\newblock Accessed: 2019-09-01.

\bibitem{dataset_nasdaq}
``{NASDAQ} price,''
  \url{https://finance.yahoo.com/quote/%5EIXIC/history?p=%5EIXIC},
\newblock Accessed: 2019-09-01.

\bibitem{dataset_usd}
``Trade weighted {USD}-index against the currencies of a broad group of trading
  partners from {January} 1995 till {August} 2019,''
  \url{https://fred.stlouisfed.org/series/TWEXB},
\newblock Accessed: 2019-09-01.

\bibitem{rios_tobar2019}
G.~Rios and F.~Tobar,
\newblock ``Compositionally-warped {G}aussian processes,''
\newblock {\em Neural Networks}, vol. 118, pp. 235 -- 246, 2019.

\bibitem{warped04}
E.~Snelson, Z.~Ghahramani, and C.~E. Rasmussen,
\newblock ``Warped {G}aussian processes,''
\newblock in {\em Advances in Neural Information Processing Systems 16}, pp.
  337--344. MIT Press, 2004.

\end{thebibliography}
\end{document}